# Robust field-free switching using large unconventional spin-orbit torque in an all-van der Waals heterostructure

Yiyang Zhang[1], Xiaolin Ren[1], Ruizi Liu[2], Zehan Chen[1], Xuezhao Wu[2], Jie Pang[3], Wei Wang[4], Guibin Lan[3], Kenji Watanabe[5], Takashi Taniguchi[6], Youguo Shi[3], Guoqiang Yu[3], Qiming Shao[#1,2]

1 Department of Physics, The Hong Kong University of Science and Technology, Hong Kong SAR
2 Department of Electronic and Computer Engineering, The Hong Kong University of Science and Technology, Hong Kong SAR
3 Beijing National Laboratory for Condensed Matter Physics, Institute of Physics, Chinese Academy of Sciences, Beijing 100190, China
4 Key Laboratory of Flexible Electronics (KLoFE) & Institute of Advanced Materials (IAM), School of Flexible Electronics (Future Technologies), Nanjing Tech University (NanjingTech), Nanjing 211816, China
5 Research Center for Electronic and Optical Materials, National Institute for Materials Science, 1-1 Namiki, Tsukuba 305-0044, Japan
6 Research Center for Materials Nanoarchitectonics, National Institute for Materials Science, 1-1 Namiki, Tsukuba 305-0044, Japan
# Corresponding emails: eeqshao@ust.hk

**The emerging all-van der Waals (vdW) magnetic heterostructure provides a new platform to control the magnetization by the electric field beyond the traditional spintronics devices. One promising strategy is using unconventional spin-orbit torque (SOT) exerted by the out-of-plane polarized spin current to enable deterministic magnetization switching and enhance the switching efficiency. However, in all-vdW heterostructures, large unconventional SOT remains elusive and the robustness of the field-free switching against external magnetic field hasn't been examined, which hinder further applications. Here we demonstrate the field-free switching in an all-vdW heterostructure combining a type-II Weyl semimetal $TaIrTe_4$ and above-room-temperature ferromagnet $Fe_3GaTe_2$. The fully field-free switching can be achieved at $2.56\times10^{10}$ A/m² at 300K and a large SOT effective field efficiency of the out-of-plane polarized spin current generated by $TaIrTe_4$ is determined to be 0.37. Moreover, we find that the switching polarity cannot be changed until the external in-plane magnetic field reaches 252mT, indicating a robust switching against the magnetic field. The numerical simulation suggests the large unconventional SOT reduces the switching current density and enhances the robustness of the switching. Our work shows that all-vdW heterostructures are promising candidates for future highly efficient and stable SOT-based devices.**

## 1. Introduction

Since the discovery of long-range ferromagnetic order in two-dimensional (2D) materials[1–3], multiple methods have been used to control the magnetization by the electric field[3–8]. Among those methods, using the spin current generated by one layer to exert spin-orbit torque (SOT) on the adjacent ferromagnetic layer can effectively switch the magnetization[9,10]. Recently developed all-van der Waals (vdW) heterostructures provide new platforms to explore the advanced application[11,12], and some advantages have been shown in all-vdW devices. In addition to the atomically flat interface[13–15], some vdW ferromagnets has intrinsically strong perpendicular magnetic anisotropy (PMA) even down to monolayer[3,16,17], which may enable dense and stable storage. Moreover, certain transition metal dichalcogenides (TMD) have exotic spin-momentum locking effects[18,19] and large spin-orbit coupling[20,21], which make them promising spin-generation layers. Particularly, low-symmetry TMD provides out-of-plane polarized spin current[22–24]. It enables magnetization switching without an external magnetic field by overcoming the intrinsic magnetic damping and enhances the switching efficiency[25,26].

Recently discovered vdW metallic ferromagnet $Fe_3GaTe_2$ maintains strong PMA above room temperature[16,17] and a few reports have used Pt and $WTe_2$ to achieve the magnetization switching[27–30]. The vdW type-II Weyl semimetal $TaIrTe_4$ has a space group 31 ($Pmn2_1$), and the crystal structure is similar to the $T_d$-phase $WTe_2$ and $MoTe_2$[31,32]. The glide plane n and the screw axis $2_1$ symmetry are broken near the surface of $TaIrTe_4$, where only a mirror plane m is preserved. When a current is injected perpendicularly to the mirror plane, the low crystal symmetry allows an out-of-plane polarized spin current to be generated. Recently, a few experiments revealed the charge-to-spin conversion in $TaIrTe_4$ is higher than $WTe_2$[33,34,56]. Due to the larger spin Hall conductivity of $TaIrTe_4$ and atomically flat interface in the vdW heterostructure, more robust switching with higher switching efficiency should be expected in $TaIrTe_4/Fe_3GaTe_2$. However, the strength of the unconventional SOT exerted by the out-of-plane polarized spin current in all-vdW heterostructures hasn't been quantified so far and the robustness of switching against the magnetic field hasn't been examined[29,35-37]. These may hinder the further application of using all-vdW heterostructure for spintronics devices.

Here we report the realization of the field-free switching in the all-vdW heterostructure $TaIrTe_4/Fe_3GaTe_2$ using the unconventional SOT exerted by few-layer $TaIrTe_4$. The fully magnetization switching can be achieved at $2.56\times10^{10}$ A/m$^2$ at 300K with a high switching energy efficiency. The effective field induced by the unconventional SOT is around $2.2$mT/($1\times10^{10}$ A/m$^2$) determined by the loop-shift method, which can be converted to the SOT effective field efficiency as 0.37. To verify the robustness of the switching against the external magnetic field, we also impose an in-plane magnetic field during the switching process. We find that the switching polarity cannot be changed until the magnetic field reaches 252mT, and this large critical magnetic field indicates a stable magnetization switching. Both unconventional SOT effective field and critical magnetic field are one order larger than previously reported $TaIrTe_4/CoFeB$[33,34]. According to our numerical simulation result, we attribute the high switching efficiency and the stability against magnetic field to the large unconventional SOT that exerted on $Fe_3GaTe_2$. Our work suggests that all-vdW heterostructure can be considered as a promising candidate for the future spintronics device.

## 2. Result

In this study, we fabricate $TaIrTe_4/Fe_3GaTe_2$ devices with covered hexagonal boron nitride (hBN) pieces to protect these devices from air-degradation (see Methods). The 6nm platinum electrode is used to reduce the effect of strain. hBN, few-layer $TaIrTe_4$, and $Fe_3GaTe_2$ are mechanically exfoliated on the silicon wafer and then transferred to the electrode. The schematic is shown in the left part of Fig. 1a. The right part of Fig. 1a shows the crystal structure of $TaIrTe_4$ and $Fe_3GaTe_2$. When the current J flows perpendicularly to the mirror plane m (along a-axis), the out-of-plane polarized spin current will be generated in addition to the conventional in-plane polarized spin current. Thus, the polarization direction will tilt to the out-of-plane direction with an angle of $\psi$, which is defined as spin canting angle. Fig. 1b shows the optical image of device D1. The a-axis of TaIrTe$_4$ can be determined to be along the straight long side of the exfoliated flake. It is further confirmed by the angle-resolved polarized Raman spectroscopy (see Supplementary Note 1). The thickness of TaIrTe$_4$ and Fe$_3$GaTe$_2$ in device D1 are 10.3nm and 4.8nm from the atom force microscope results.

The magnetotransport measurement results are shown in Figs. 1c-1e (see Methods). When the magnetic field is perpendicular to the sample surface (along c-axis), a square anomalous Hall loop can be obtained from 180K to 320K, as shown in Fig. 1c. The coercivity B$_c$ is 16mT and the anomalous Hall resistance is 4Ω at 300K. At 340K, the square loop disappears, suggesting a Curie temperature above room temperature. When the magnetic field sweeps parallel with the sample surface, a parabolic shape appears in the low field range and the curve gradually becomes flat at the high field (Fig. 1d), which indicates a strong effective PMA. The effective PMA field can be extracted by fitting the parabolic part, and the PMA energy density K$_u$ can be calculated by adopting the saturation magnetization M$_s$ from ref. 28. B$_c$ and K$_u$ decrease with increasing temperature, as shown in Fig. 1e. The PMA energy density of Fe$_3$GaTe$_2$ is around $4.3 \times 10^5$ J/m$^3$, which is consistent with the previous report[16].

To verify the existence of unconventional SOT and quantify the strength, we do the loop-shift measurement without an in-plane magnetic field. To reduce the Joule heating at the large current (see Supplementary Note 2), a short pulse current (500 $\mu$s) is imposed to measure the anomalous Hall loop. When the current flows along a-axis, the out-of-plane polarized spin current will exert unconventional SOT on Fe$_3$GaTe$_2$. If the unconventional SOT is large enough to overcome the intrinsic magnetic damping, a shift from the original point at the center of hysteresis loop will appear. The SOT effective field at a certain current can be extracted as $B_{eff} = (B_{center}^+ - B_{center}^-)/2$, where $B_{center}^+$ ($B_{center}^-$) represents the center of the anomalous Hall loops when positive (minus) current is imposed.

In the Fe$_3$GaTe$_2$/TaIrTe$_4$ device D1, the hysteresis loops are identical when a low current ($\pm$1mA) is injected along a-axis. While for a large current ($\pm$3mA), two AHE loops are well separated from each other, proving the existence of the unconventional SOT (Fig. 2a). Note that we also see a large difference in coercive field B$_c$ when we apply positive and negative current (see Supplementary Note 3). By contrast, when we impose current along b-axis, we observe no clear shift, as shown in Fig. 2b. To determine the strength of the unconventional SOT, we measure the current dependence of the SOT effective field, and three sets of AHE loops are collected at every current. The average value of the SOT effective field is calculated and plotted in Fig. 2c. In the high current range, the SOT effective field can be fitted by a straight line and the slope represents the SOT effective field per current[33,34,38]. Taking the conductivity of Fe$_3$GaTe$_2$ and TaIrTe$_4$ as $4.3 \times 10^5 \, \Omega^{-1} m^{-1}$ and $5.5 \times 10^5 \, \Omega^{-1} m^{-1}$ (a-axis), we extract SOT strength as 2.2mT/($1 \times 10^{10}$ A/m$^2$). The SOT effective field per current can be further converted to the SOT effective field efficiency as 0.37, according to the equation $\xi_z^{B_{eff}} = 2eM_s t B_{eff}/\hbar J$. Here, $\xi_z^{B_{eff}}$ represents the unconventional SOT effective field efficiency, $M_s$ is the saturation magnetization (here we choose $M_s$ to be $1.7 \times 10^5$ A/m[28]), $t$ is the thickness of the ferromagnet layer, $B_{eff}$ is the SOT effective field, and $J$ is the electric current density. Note the SOT effective field efficiency defined here may deviate from the damping-like SOT efficiency determined from the second harmonic Hall measurement or spin torque-ferromagnetic resonance (ST-FMR), where the damping-like SOT field can be measured in the single magnetic domain. We cannot perform second harmonic or ST-FMR measurements to extract the damping-like SOT efficiency because of very small planar Hall resistance (or anisotropic magnetoresistance) and large PMA (see Supplementary Note 9). We also discuss the relation between the damping-like SOT efficiency and the slope extracted from the loop-shift method using numerical simulations (see Supplementary Note 8). On the other hand, when measured along b-axis, the slope is at least one order smaller than the result along a-axis, indicating a negligibly small contribution from unconventional SOT. Notice that due to the intrinsic damping is not compensated when the current is small, the fitted line does not cross the original point[33,34,38]. Here we summarize $\xi_z^{B_{eff}}$ and the charge conductivity $\sigma_c$ in Fig. 2d, and the comparison is limited within loop-shift results to avoid errors introduced by different measurement methods. We also characterize the conventional SOT induced by the in-plane polarized spin current by measuring the loop-shift with in-plane external magnetic field[38]. The conventional SOT effective field is extracted by measuring the shift of AHE loops under positive and minus currents while fixing the magnitude of the current. The SOT effective field efficiency $\xi_y^{B_{eff}}$ is determined to be 0.15-0.16 along b-axis, which are similar to the reported values in references[33,34]. All results of loop-shift measurements for $\xi_z^{B_{eff}}$ and $\xi_y^{B_{eff}}$ in three devices are shown in Supplementary Notes 4-7 and Supplementary Table S1.

Using the large unconventional SOT, the field-free switching can be achieved above room temperature. The deterministic switching can be clearly observed when the current is applied along a-axis, and the full switching can be observed at I=±3mA. On the contrary, when I // b-axis, only in-plane spin polarization can be generated, and no switching can be achieved without the assistance of an external magnetic field. As shown in Fig. 3a, the m=0 final state when I // b-axis indicates a random distribution of m=1 and m=−1 domain in $Fe_3GaTe_2$. To examine the repeatability of the field-free switching, we use multiple pulses I=±3.4mA to switch the magnetization, and the device shows a highly repeatable high or low Hall resistance according to the positive or minus current (Fig. 3b). To test the robustness of the switching against the temperature variation, we measure the magnetization switching at different temperatures. The full switching can be achieved at the temperatures from 285K to 305K (Fig. 3c), and the critical current density $J_c$ decreases as the temperature rises (Fig. 3d). Since the low temperature will enlarge the PMA effective field and saturation magnetization, it leads to a higher $J_c$. At 300K, $J_c$ is 2.56×10$^{10}$ A/m$^2$, and the switching efficiency can be calculated as 1.49 according to the equation $\eta = 2eM_s tB_c/\hbar J_c$ [40,41], when saturation magnetization $M_s$ is taken as 1.7×10$^5$ A/m[28].

To show the robustness of switching against the external magnetic field, we also do the switching measurement under an in-plane magnetic field. For magnetization switching induced purely by the in-plane polarized spin current, the switching polarity depends on the direction of the in-plane magnetic field and the switching-polarity-changed critical field $B_{critical}$ is 0. On the other hand, in the presence of the out-of-plane polarized spin current, $B_{critical}$ shifts from 0 since the switching by the out-of-plane polarized spin current doesn't change its polarity under the magnetic field. As shown in Fig. 3e, the switching behavior distinguishes under negative and positive magnetic fields. The switching amplitude becomes zero at 252mT and the polarity changes if the magnetic field further increases. By contrast, the polarity doesn't change under the negative magnetic field. We define the switching ratio, $\Delta R_{xy}(B)/\Delta R_{xy}(0)$, where $\Delta R_{xy}(B)$ represents the Hall resistance change in the switching from large negative to large positive current under the in-plane magnetic field $B$. The switching ratio changes its sign when the switching polarity changes. The switching ratio as a function of the external magnetic field is plotted in Fig. 3f. It is worth noticing that the critical field in TaIrTe$_4$/Fe$_3$GaTe$_2$ is more than one order larger than TaIrTe$_4$/CoFeB[33,34].

To explore the origin of the robustness, we perform the numerical macrospin simulation based on Landau-Lifshitz-Gilbert (LLG) equation. We take the thermal effect into consideration by adding the thermal effective field (see Method)[42]. The damping-like SOT term in LLG equation is proportional to $(\boldsymbol{m} \times \boldsymbol{\sigma}) \times \boldsymbol{m}$, where $\boldsymbol{m}$ and $\boldsymbol{\sigma}$ is the unit vector of magnetization and spin polarization, respectively. Here we only include the anti-damping term as the results capture the essential physics as we show below. In heavy metal or topological insulator/ferromagnet heterostructure, $\boldsymbol{\sigma}$ is taken as in-plane direction. In the TaIrTe$_4$/Fe$_3$GaTe$_2$, $\boldsymbol{\sigma}$ tilts to the out-of-plane direction with a spin canting angle $\psi$. First, we simulate the relation between the spin canting angle and the switching ratio (Fig. 4a). For every $\psi$ and external magnetic field, 800 times switching simulations for a certain current are done, and the final magnetization M is taken as the average value of these simulations. By sweeping the current, a switching loop is calculated, so the switching ratio can be obtained. For $\psi = 0°$, the critical magnetic field $B_{critical}$ is 0 since the switching polarity changes at $B_x = 0$. The zero-switching ratio at $B_x = 0$ means that field-free deterministic switching cannot be achieved without the unconventional SOT. As $\psi$ increases, the critical field gets larger and the switching current $J_c$ decreases (Fig. 4b). The numerical simulation results suggest that the higher switching efficiency and larger polarity-change critical field in TaIrTe$_4$/Fe$_3$GaTe$_2$, compared to TaIrTe$_4$/CoFeB, can be attributed to larger unconventional SOT that exerted on Fe$_3$GaTe$_2$. The experimental switching efficiency and the critical magnetic field are summarized in Fig. 4c. We see that our device achieves the record-high robustness against the in-plane field disturbance while maintaining a high energy efficiency.

## 3. Discussion and conclusion

We show the field-free switching in the all-vdW heterostructure TaIrTe$_4$/Fe$_3$GaTe$_2$ at and above room temperature with high switching efficiency. The strengths of unconventional SOT and switching-polarity-change magnetic field are further determined. The high experimental critical magnetic field and switching efficiency can be attributed to the large unconventional SOT, according to our numerical simulation results. Notice that the critical magnetic field and unconventional SOT effective field efficiency are both one order larger than the heterostructure consisting of TaIrTe$_4$ and a non-vdW ferromagnet CoFeB[33,34]. Here we propose several possible reasons for this. First, compared to the magnetron sputtering method, the dry-transfer method keeps the surface of TaIrTe$_4$ high-quality[43]. As the space group of bulk TaIrTe$_4$ is Pmn2$_1$, which forbids the generation of out-of-plane polarized spin current[44,45], the out-of-plane polarized spin current is mainly generated near the surface of TaIrTe$_4$. As a result, the unconventional spin conversion highly depends on the surface quality. The dry-transfer method will not do damage to the surface of materials, which may lead to a large unconventional spin generation. Second, the interfacial magnetic spin Hall effect has been discovered in the MoTe$_2$/Fe$_3$GeTe$_2$[46], where the misalignment between the mirror planes of MoTe$_2$ and Fe$_3$GeTe$_2$ breaks the interfacial symmetry and leads to the generation of T-odd spin current. Since MoTe$_2$ has the similar crystal structure to TaIrTe$_4$, and Fe$_3$GaTe$_2$ is the counterpart of Fe$_3$GeTe$_2$, the similar spin-generation mechanism should also exist in TaIrTe$_4$/Fe$_3$GaTe$_2$. Third, the strength of spin torque exerted on the ferromagnet layer is not only determined by the spin source material but also dependent on the electronic band structure of the ferromagnet[47,48]. Since the few-layer Fe$_3$GaTe$_2$ is a highly anisotropic material compared to CoFeB, the response to the spin current in Fe$_3$GaTe$_2$ and CoFeB may be different. Finally, the atomically flat interface may also enlarge the spin transparency. These possible reasons are beyond the scope of our experiment, and further theory and experiment investigations remain to be explored.

In summary, we show a highly efficient field-free switching in all-vdW TaIrTe$_4$/Fe$_3$GaTe$_2$. The unconventional SOT and polarity-change critical magnetic field are found to be quite high, compared to TaIrTe$_4$/CoFeB. The efficient and robust magnetization switching show the advantage of using all-vdW heterostructure as the next generation spintronics device, and many possibilities like gate-controlled spin generation in all-vdW heterostructures are waiting to be explored. For the industrial application, with the success of growing wafer-scale vdW ferromagnets with controlled thickness[49,50,15], integrating them into spintronics device for industrial application should also attract more attention.

**Figures and captions**

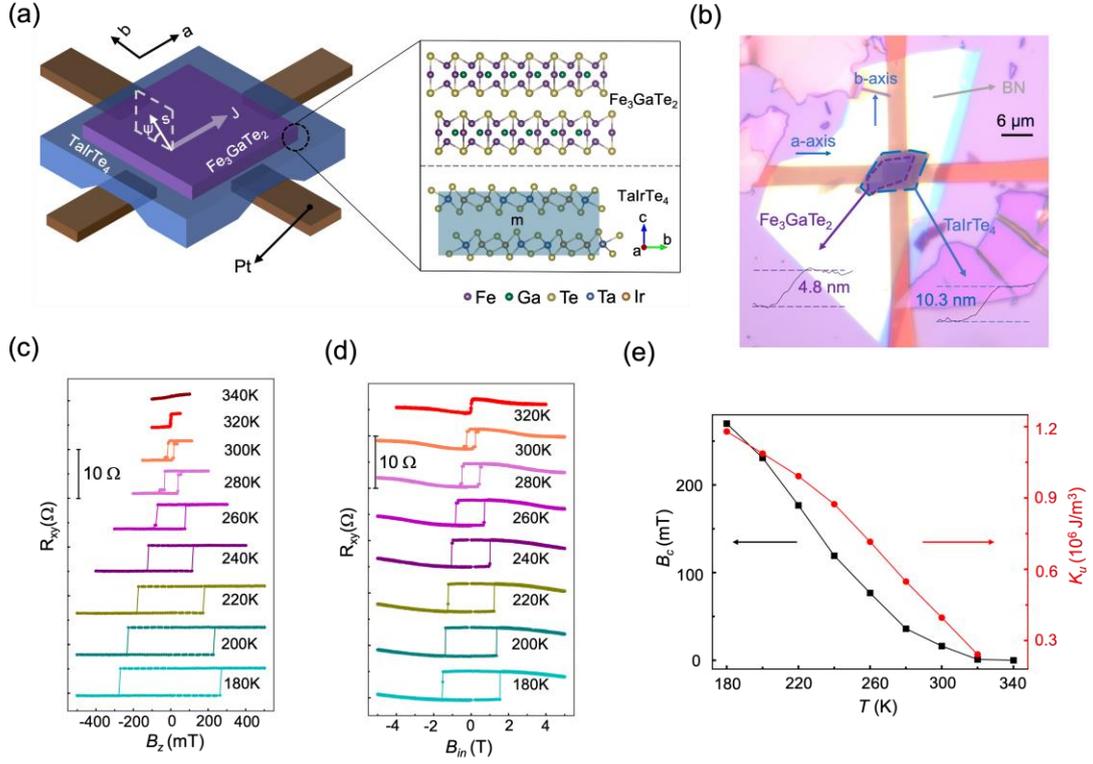

**Figure 1.** Schematic diagram and magnetic characterization of TaIrTe$_4$/Fe$_3$GaTe$_2$ heterostructure D1. a) Schematic diagram of TaIrTe$_4$/Fe$_3$GaTe$_2$ heterostructure. The right part shows the crystal structure of TaIrTe$_4$ and Fe$_3$GaTe$_2$. The inset shows the a, b and c axis of TaIrTe$_4$ and the blue square m is the mirror plane of the few-layer TaIrTe$_4$, which is perpendicular to a-axis. When current flows along a-axis, a spin current with polarization $s$ which tilts from the in-plane direction as $\psi$ angle will be generated. b) Optical image and the atom force microscope result of device D1. The a(b)-axis is indicated by the blue arrows in the picture. The thickness of TaIrTe$_4$ and Fe$_3$GaTe$_2$ determined from AFM is shown in the figure. c,d) Hall resistance as a function of perpendicular magnetic field and in-plane magnetic field at different temperatures. Here an offset is added on R$_{xy}$ to separate different Hall resistance loops. e) Extracted coercivity field and magnetic anisotropy energy density at different temperatures.

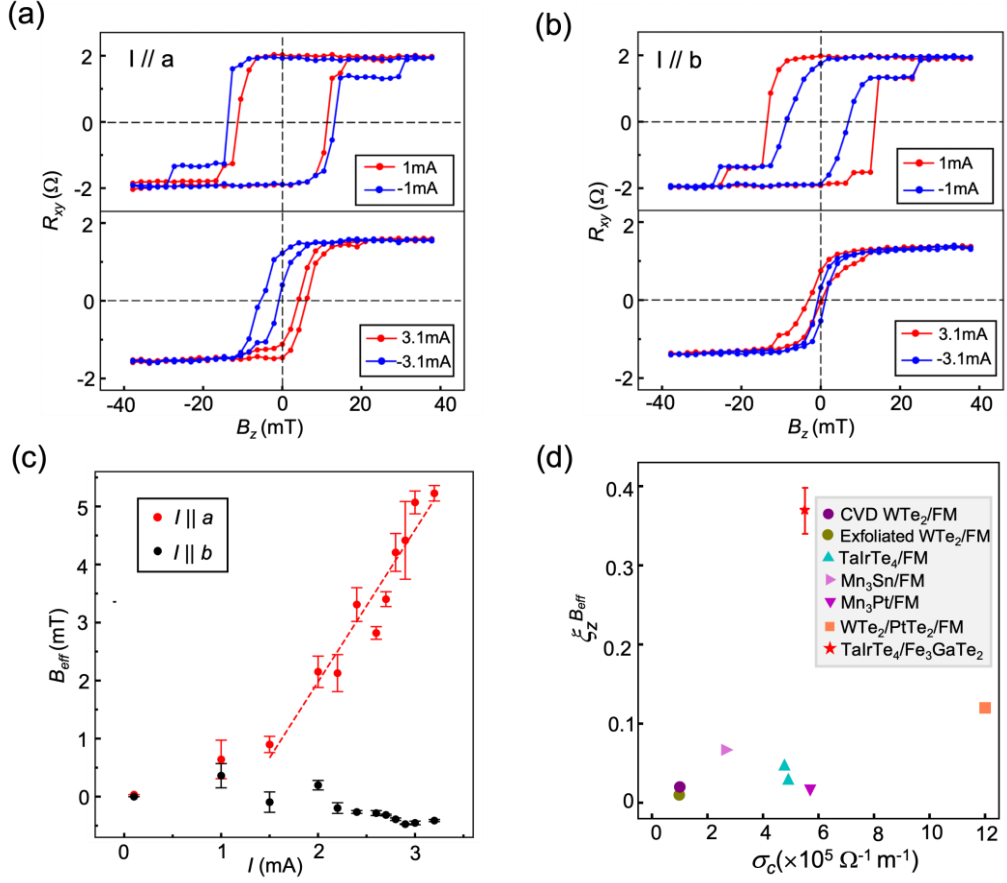

**Figure 2.** Loop-shift result and the summary of unconventional SOT effective field efficiency, charge conductivity. a, b) Anomalous Hall loop under 1mA and 3.1mA when current is imposed along a-axis or b-axis. The centers of the anomalous Hall loops shift at $I = \pm 3.1$mA when $I//a$ suggests the existence of the unconventional SOT. c) The SOT-induced effective field as the function of current along a-axis and b-axis. When $I//a$, The effective field at large currents can be well fitted by the red dash line and the unconventional SOT effective field efficiency can be extracted. While for $I//b$, the nearly zero $B_{eff}$ shows the absence of unconventional SOT. The error bar is the standard derivation of three repeated measurements. d) The summary unconventional SOT effective field efficiency and charge conductivity. FM represents the traditional ferromagnetic layer, including CoFeB, Pt/Co/Pt, etc. Benchmark data are from references[24,33,34,38,51–53]. The error bar is the standard derivation of three repeated measurements.

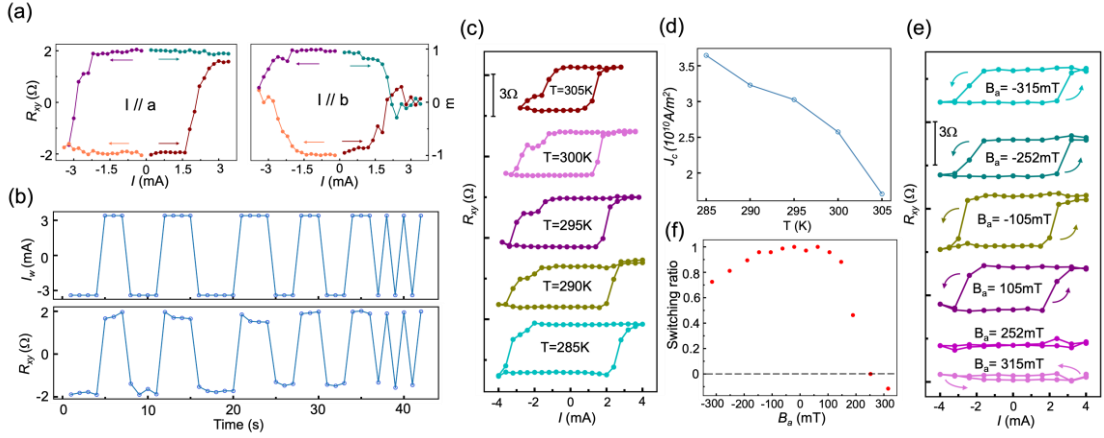

**Figure 3.** Field-free switching of TaIrTe$_4$/Fe$_3$GaTe$_2$ heterostructure. a) Deterministic switching can(cannot) be achieved when current is imposed along the a(b)-axis. b) Repeatable switching using ± 3.4mA current, applied along a-axis. c) Switching loops at different temperature. d) Temperature dependence of the critical switching current. The decrease on the switching current density as temperature rises suggests the assistance from thermal effect in the switching process. e) Switching loops at different magnetic field when I // a-axis. $B_a$ represents that the magnetic field is along the a-axis. The arrows on the switching curve indicate the switching polarity. f) Variation of the switching ratio, $\Delta R_{xy}(B)/\Delta R_{xy}(0)$, with the external magnetic field $B$. Here, $\Delta R_{xy}(B)$ represents the Hall resistance change in the switching from large negative to large positive current under the in-plane magnetic field $B$. The polarity-changed magnetic field is 252mT, which is record-high at room temperature.

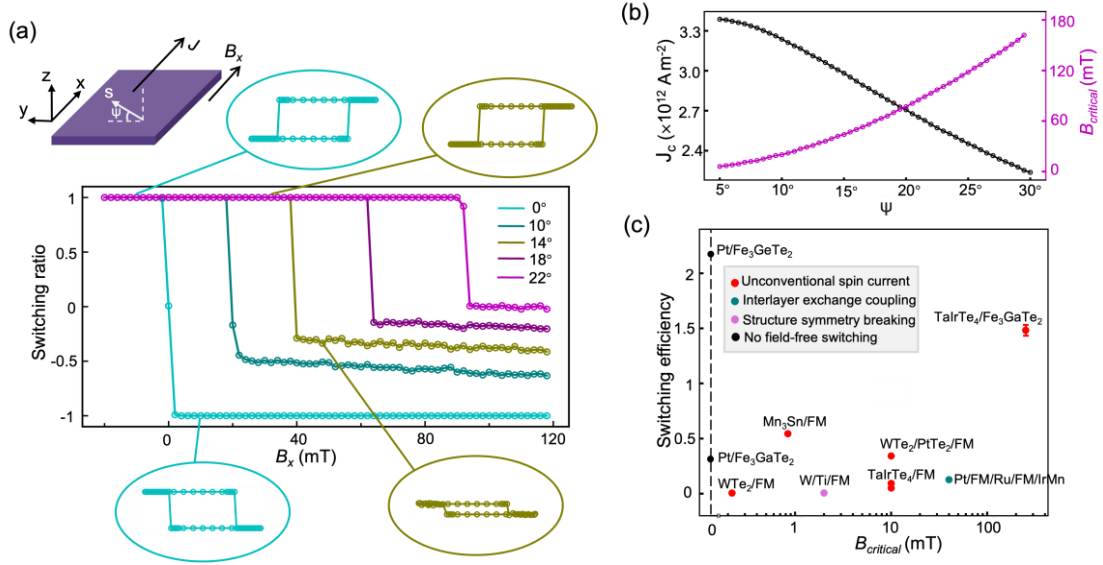

**Figure 4.** Numerical simulations and the benchmark for the switching efficiency and critical magnetic field. a) The switching ratio changes with the external magnetic field for various spin canting angles. The upper left part shows the definition of x, y and z-axis. The spin canting angle is defined as the angle between the spin polarization and sample plane. Four insets show simulated current-induced switching loops for different spin canting angles. b) The relation of switching critical current and critical magnetic field of switching polarity reversal with the spin canting angle in the simulation results. c) The benchmark of experimental critical magnetic field and switching efficiency. Benchmark data are from references[28,33,34,38,40,51,52,54,55]. The error bar is the standard derivation of seven repeated measurements.

## 4. Methods

*Crystal Growth*: The growth method of TaIrTe$_4$ has been described in the previous published procedures[34]. Fe$_3$GaTe$_2$ crystals were from two source. One was purchased from Nanjing MKNANO Tech. Co., Ltd. The other was grown via a self-flux method. Fe, Ga, and Te were used in a molar ratio of 1:1:2, with high-purity Fe powder (99.99%), Ga lumps (99.9999%), and Te powder (99.999%), placed in a vacuum quartz tube and sealed. The mixture was first heated to 1000°C within 6 hours and maintained at 1000°C for 24 hours. Then, it was cooled down to 880°C within 1 hour, followed by a week-long cooling to 780°C, and maintained at 780°C for 2 days. After reaching room temperature, the sample was crushed, and Fe$_3$GaTe$_2$ single crystals were separated using a magnet.

*Device Fabrication*: Bulk TaIrTe$_4$ and Fe$_3$GaTe$_2$ were mechanically exfoliated on Si/SiO2(285nm) on the N$_2$-filled glovebox. The flakes with appropriate thickness were then optically selected. A PDMS/PC stamp was used to pick up hBN, Fe$_3$GaTe$_2$ and TaIrTe$_4$ in sequence. To prepare the electrodes, the electron beam lithography was used to define the inner pattern with PMMA. The magnetron sputtering was then carried out to grow 6nm platinum as the electrode. After that, the second electron beam lithography combined with electron beam deposition were used to make the outer pattern Ti(5nm)/Au(50nm).

*Transport Measurement*: The room-temperature transport measurement is conducted in a custom-built measurement stage and the temperature-varied measurement is done in the Cryogen Free Measurement System (CFMS). The Keithley 2440 and the Kepco current source are used to offer the current for magnet in the custom-built measurement stage. The Keithley 6221 current source and the Keithley 2182 nanovoltmeter are used to measure the loop-shift curve and magnetization switching.

*Numerical Simulation:* The numerical macrospin simulation is carried out based on the LLG equation combined with the thermal field as

$$\frac{d\boldsymbol{m}}{dt} = -\gamma \boldsymbol{m} \times \boldsymbol{B}_k + \alpha \boldsymbol{m} \times \frac{d\boldsymbol{m}}{dt} + \gamma B_{SOT}[(\boldsymbol{m} \times \boldsymbol{\sigma}) \times \boldsymbol{m}] + \boldsymbol{B}_{thermal}$$

$\gamma$ is the gyromagnetic ratio, $\alpha$ is the Gilbert damping constant, $\boldsymbol{m}$ and $\boldsymbol{\sigma}$ is the unit vector of magnetization and spin polarization, $\boldsymbol{B}_k$ is the magnetic anisotropy field, $B_{SOT}$ is the magnitude of damping-like SOT field, $\boldsymbol{B}_{thermal}$ is the thermal field. Here, $\boldsymbol{B}_k = (2K_u/M_s)m_z\hat{z}$, $\boldsymbol{\sigma} = (0, cos\psi, -sin\psi)$ ($\psi$ is the spin canting angle), $B_{SOT} = \hbar\theta J/2et_{FM}M_s$ ($\theta$ is the spin hall angle, $J$ is the electric current density, $t_{FM}$ is the thickness of the ferromagnetic layer, $M_s$ is the saturation magnetization), $B_{thermal} = \boldsymbol{g}\sqrt{2k_BT\alpha/M_s\gamma V\Delta t}$ ($T$ is temperature, $V$ is the volume of the ferromagnetic layer, $\Delta t$ is the time-step of the magnetization evolution, $\boldsymbol{g}$ is the unit vector consist of three independent Gaussian random variables). We set this parameter as: $M_s = 2.2 \times 10^5$ A/m, $K_u = 1 \times 10^5$ J/m$^3$, $\alpha = 0.2$, $\theta = 0.2$, $t_{FM} = 3$ nm, $T = 300$K, $V = 9 \times 10^{-21}$ m$^3$. The simulation time-step is 1 ps. We simulate the magnetization switching 800 times for every current to get 800 final states and the magnetization is taken as the average value of these final states to generate a switching loop. This averaging is simulating the multi-domain effect in the micrometer-size TaIrTe$_4$/CoFeB devices.

## Acknowledgement

The authors thank Zhencun Pan, Zhiming Liao, Xingkai Cheng, Junwei Liu, Chen Liu, Xixiang Zhang, Yi Wan, Lain-Jong Li for inspiring discussions. The authors at HKUST acknowledge funding support from NSFC/RGC Joint Research Scheme (No. N_HKUST620/21), National Key R&D Program of China (Grants No.2021YFA1401500), Research Grant Council—Early Career Scheme (Grant No. 26200520), General Research Fund (Grant No. 16303521), and the State Key Laboratory of Advanced Displays and Optoelectronics Technologies. The authors at IOP, CAS are supported by the National Natural Science Foundation of China (Grant No. 52161160334). K.W. and T.T. acknowledge support from the JSPS KAKENHI (Grant Numbers 21H05233 and 23H02052) and World Premier International Research Center Initiative (WPI), MEXT, Japan.

## Conflicts of interest

The authors declare that they have no conflict of interest.

## Author Contributions

Y. Z. fabricated the devices, performed the measurement and analyzed the data with assistance from X. R. and R. L. X. R. and Y. Z. performed the Raman spectroscopy measurement. Z. C. performed the numerical simulation. X. W. helped to fabricate the platinum electrode. G. L., G. Y., J. P. and Y. S offered the TaIrTe$_4$ crystal and W. W. grew the Fe$_3$GaTe$_2$ crystal. K. W., T. T. offer the high-quality hBN. Q. S. conceived and supervised the project. Y. Z. wrote the manuscript, and all authors discussed the content of the manuscript.

## Data Availability Statement

The data that support the findings of this study are available from the corresponding author upon reasonable request.